\documentclass{PoS}

\title{Correlations in the associative production of $B_c$ and $D$ mesons at LHC}

\ShortTitle{Correlations in the associative production of $B_c$ and $D$ mesons at LHC}

\author{\speaker{Aleksander Berezhnoy}%
         \thanks{Authors thank E.E.~Boos, L.K.~Gladilin, N.V.~Nikitin  and S.P.~Baranov for the fruitful discussion.
The work of A.V.~Berezhnoy was partially supported by Federal Agency for Science and Innovation under state contract 02.740.11.0244.
}\\
        SINP MSU, Moscow, Russia\\
        E-mail: \email{aber@trtk.ru}}

\author{Anatolii Likhoded\\
        IHEP, Provtino, Russia\\
        E-mail: \email{Anatolii.Likhoded@ihep.ru}}

\abstract{It is shown that the study of correlations
  in the associative production of $B_c$ and $\bar D$ mesons at LHC  
allows to obtain the essential  information about the $B_c$ production mechanism.
}

\FullConference{The XIXth International Workshop on High Energy Physics and Quantum Field Theory \\
		 8-15 September 2010\\
		 Golitsyno, Moscow, Russia}

\begin{document}

\section{Introduction}
Recent measurements of $B_{c}$ meson mass and lifetime in CDF \cite{Bc_CDF}
and D0 \cite{Bc_D0} experiments  are the first steps in the experimental research of quarkonia with open flavor. The measurement results  are in a good agreement with the theoretical predictions for the $B_c$ mass~\cite{Bc_spectroscopy_Kiselev, Bc_spectroscopy_Eichten, Bc_spectroscopy_Galkin}:
$$m_{B_c}^{\rm CDF}=6.2756\pm 0.0029(\textrm{stat.})
\pm 0.0025(\textrm{sys.}) \;\textrm{GeV}, $$
$$m_{B_c}^{\rm D0}=6.3000\pm 0.0014(\textrm{stat.})
\pm 0.0005(\textrm{sys.}) \;\textrm{GeV},
$$
$$ \qquad  m_{B_c}^{\rm theor}=6.25\pm 0.03 \;\textrm{GeV};$$
as well as for the decay time~\cite{Bc_decay_SR_Kiselev,Bc_decay_PM_Kiselev}:
$$\tau_{B_c}^{\rm CDF}=0.448^{+0.038}_{-0.036} (\rm stat.)\pm 0.032 (\rm sys.)\;\textrm{ps},$$
$$\tau_{B_c}^{\rm D0}=0.475^{+0.053}_{-0.049} (\rm stat.)\pm 0.018 (\rm sys.)\;\textrm{ps},$$
$$\tau_{B_c}^{\rm theor}=0.48 \pm 0.05 \;\textrm{ps}.$$

Unfortunately, the experimental estimations of the cross section value were not published. 
Thus, the  mechanism of $B_c$ meson production can not be understood from the obtained data due to poor experimental statistics, as well as due to large uncertainties in the theoretical predictions.
Only the planed experimental research at LHC, where about $10^{10}$ events with $B_c$ mesons per year are expected, will improve the situation. This huge amount of events will allow to obtain the information on the production cross section distributions, on the decay branching fractions, and in some cases, on the distributions of decay products. 
In this research we try to fill the gap in our theoretical understanding of $B_c$ production  and show that the study of correlations in  the associative production of $B_c$ and $\bar D$ meson at LHC  
could be an essential information source of $B_c$ production mechanism. Here we study the production characteristics which weakly depend on  parameters: the cross section distribution shapes and the ratio between $B_c^*$ and $B_c$ yields. 

\section{Fragmentation and recombination contributions into $B_c$ production}

The $B_c$ production amplitude within the discussed approach can be subdivided into two parts: the hard production of two heavy quark pairs calculated  in the 
framework of perturbative QCD and the  soft nonperturbative 
binding of $\bar b$ and $c$ quarks into quarkonium  described by nonrelativistic  wave function.
The  calculations within the discussed technique are the most simple for  the process of $B_c$  production in the  $e^+e^-$ annihilation.  As it was shown in~\cite{Bc_fragmentation_S_wave_Braaten, Bc_fragmentation_S_wave_Kiselev, Bc_production_Kolodziej_Leike_Ruckl},  the special choice of the gluonic field gauge allows to  interpret the $B_c$ production process as the $\bar b$ quark production  followed by the fragmentation of $\bar b$ quark into $B_c$ meson.
Thus,  in the $e^+e^-$ annihilation at large energies 
the consideration of leading diagrams for the $B_c$ meson production leads the well known 
factorized formula  for the  cross section distribution over $z=2 E_{B_c}/\sqrt{s}$: 
\begin{equation}
\frac{d\sigma}{dz} = \sigma_{b \bar b}\cdot D(z).
\end{equation}
  The analytical forms of fragmentation functions for $S$ wave states are known from 
\cite{Bc_fragmentation_S_wave_Braaten, Bc_fragmentation_S_wave_Kiselev}.

 The relative yield of $B_c^*$ and $B_c$ in the $e^+e^-$ annihilation obtained within pQCD calculation  $R^{B_c}_{e^+e^-}=\sigma(B_c^*)/\sigma(B_c)\sim 1.4$. Thus the naive spin counting which fairly predicts this ratio for $B^*$  and $B$ ($R^{B}_{e^+e^-}\sim 3$) can not be applied to $B_c$ and $B_c$ production.

At first sight  it would be reasonable to assume that for the gluonic $B_c$ production the fragmentation mechanism is also dominant at least from  transverse momenta larger than $B_c$ mass. But as it was shown in \cite{Bc_hadronic_production, Bc_production_Chang_Chen_Han_Jiang, Bc_production_Kolodziej_Leike_Ruckl,Bc_production_Baranov} the other mechanism essentially contribute to this process practically all over the phase space and the fragmentation approach is valid only at transverse momenta larger than $5\div 6$ masses of $B_c$.
The  total gluonic cross section predicted  using full set of leading order diagrams essentially differ from the fragmentation approach in  absolute value as well as in shape of  interaction energy dependence.

As it is predicted within pQCD~\cite{Bc_hadronic_production,Bc_production_Chang_Chen_Han_Jiang}, about 90 \% of the $B_c$ mesons at LHC energies will be produced in the gluonic fusion (Fig.~\ref{fig:diagr_gg}). Therefore in our pQCD calculations we can neglect the other partonic subprocess. 
The convolution of the gluonic subprocess cross section with the gluonic structure functions partially hides the  differences between the pQCD predictions and the fragmentation approach.
The predicted ratio between the hadronic cross section values is about of 2. Obviously, such a difference  is not  essential for the  calculations  of forth order on $\alpha_s$.
Nevertheless, the relative yield of $B_c^*$ and $B_c$ does not depend on  $\alpha_s$ and could indicate the production mechanism. Even in the kinematical region where the fragmentation model could be applied   the value of $R_{\rm hadr}=\sigma_{\rm hadr}(B_c^*)/\sigma_{\rm hadr}(B_c)$ predicted within pQCD is about 2.6 instead of 1.4 obtained within the fragmentation approach. To measure this value one need to detect $B_c^*$ meson which with unit probability decays into $B_c$ and photon.   However, it is quite difficult to detect such a process experimentally due to the  small mass
difference  between $B_c^*$ and $B_c$ mesons~\cite{Bc_spectroscopy_Kiselev}: 
$
\Delta M =M_{B_c^*}-M_{B_c}= 65 \pm 15 \; \textrm{MeV}.
$
 In laboratory system  the maximum energy of emitted photon is
$\omega_{max}=\left(\gamma+\sqrt{(\gamma^2-1)}\right)\Delta M,$
where $\gamma$ is $B_c^*$ $\gamma$ factor, and
even for $B_c^*$ with energy  $\sim 30$ GeV $\omega_{max}\sim 0.7$ GeV. Thus one can conclude that there is no certainty  that the method based on separation of $B_c^*$ from $B_c$ can be used to study the $B_c$ production mechanism. 
This is why in the next chapters we suggest another way to research the $B_c$ production mechanism.

\section{The cross section distribution on the invariant mass of $D$ and $B_c$ mesons}

\begin{figure}[!t]
\centering
\resizebox*{1.\textwidth}{!}{
\includegraphics{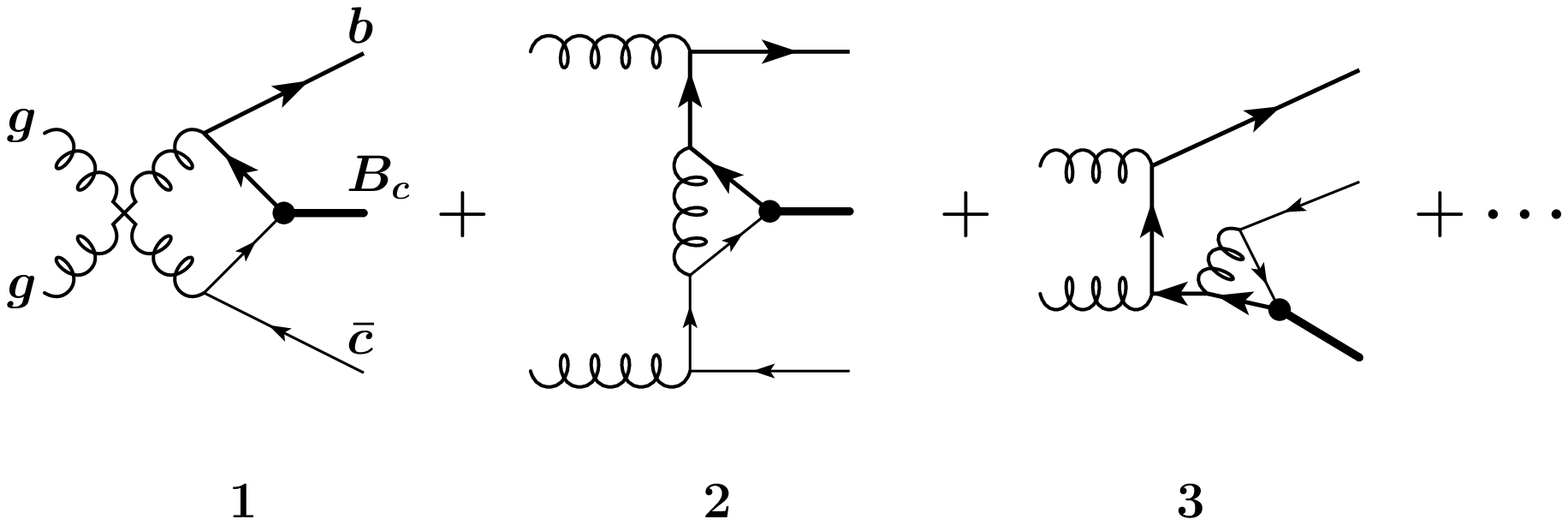}
\hfill
\includegraphics{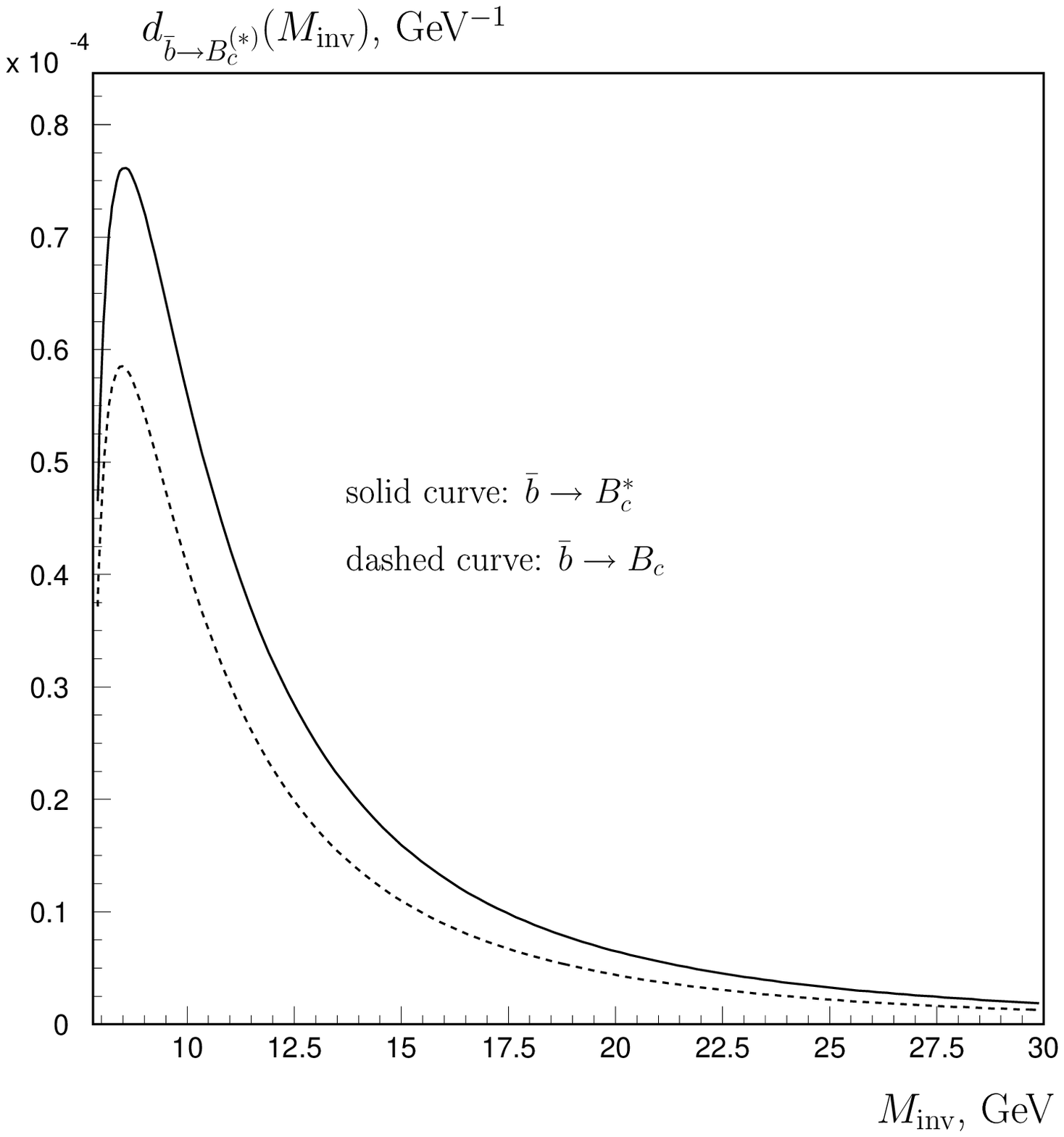}
}
\\
\parbox[t]{0.45\textwidth}{
\caption{The leading order diagrams for the process $gg\to B_c + b +\bar c$. \hfill}
\label{fig:diagr_gg}}
\hfill
\parbox[t]{0.45\textwidth}{
\caption{The normalized cross section distributions over the invariant mass of $B_c$ ($B_c^*$) and $c$-quark within fragmentation approach. \hfill}\label{fig:frag_minv}}
\end{figure}

Within fragmentation approach the shape of the cross section distribution over the invariant mass of $B_c$ and $\bar c$-quark is roughly determined by $\bar b$-quark virtuality (see the diagram (3) in Fig.~\ref{fig:diagr_gg}). 
This why the distribution should be relatively narrow. Our analytical calculations confirm this supposition~\cite{Bc_comuv}.

Here we face the problem how to transform the invariant mass of $B_c$ ($B_c^*$) and $\bar c$-quark 
$M_{B_c+\bar c}$ to invariant mass of $B_c$ ($B_c^*$) and $\bar D$-meson $M_{B_c+\bar D}$. Within the fragmentation mechanism it is naturally to assume that the $\bar D$ meson takes away the total momentum of $c$-quark, because the production of $\bar c$ quark is the last step of emission process. Such an assumption is not obvious for 
the recombination contribution. This is why two hadronization models of $\bar c$ quark have been chosen:
\begin{enumerate}
\item  $\bar D$ meson takes away the total momentum of $\bar c$-quark: $D_{c\to D}= \delta(1-z)$;
\item $\bar D$ meson takes away part of $\bar c$-quark momentum according to Kartvelishvily-Petrov-Likhoded
fragmentation function: $D_{c\to D}= z^{2.2}(1-z)$.
\end{enumerate}

\begin{figure}[!t]
\centering
\resizebox*{1.\textwidth}{!}{
\includegraphics{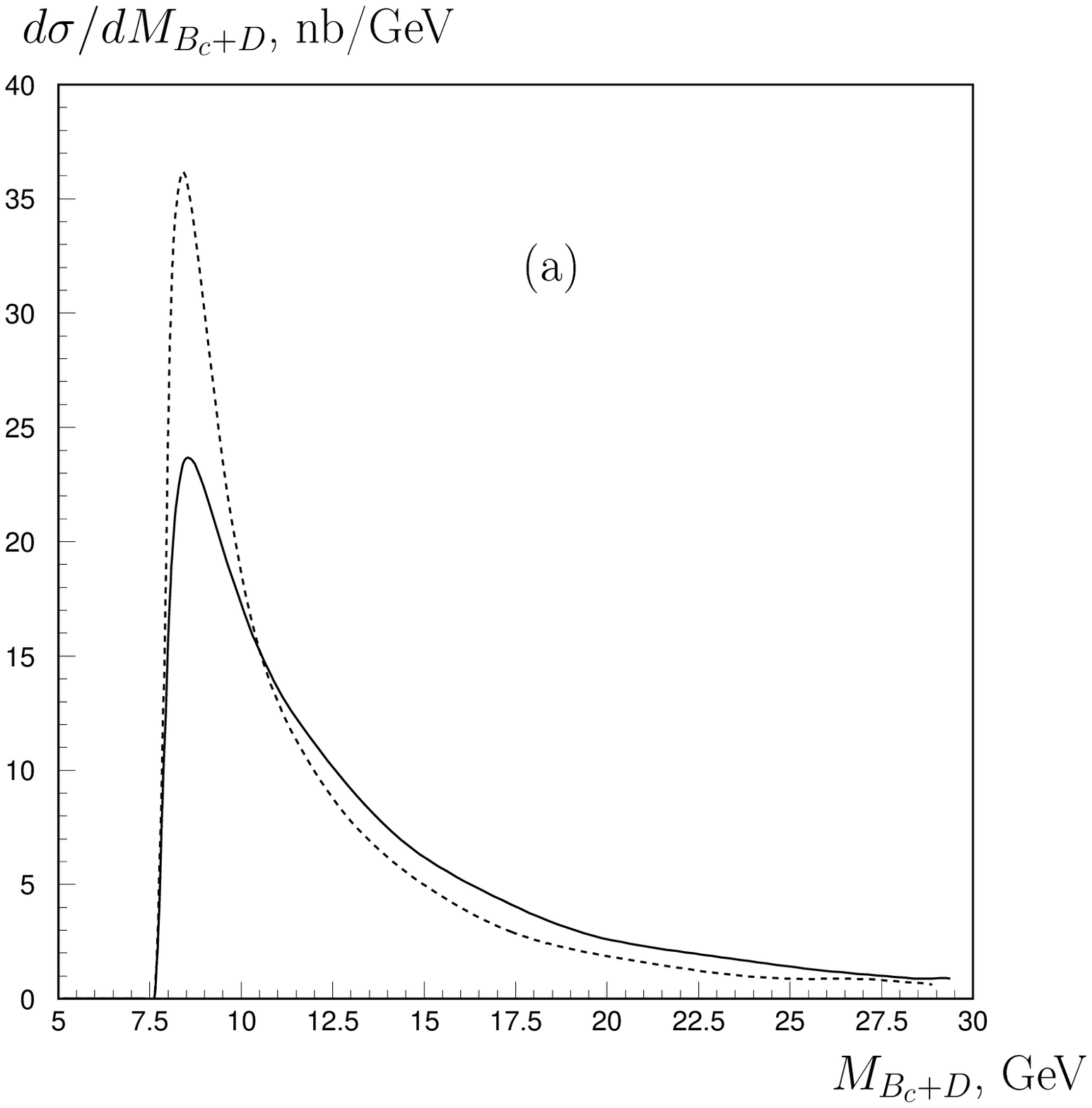}
\hfill
\includegraphics{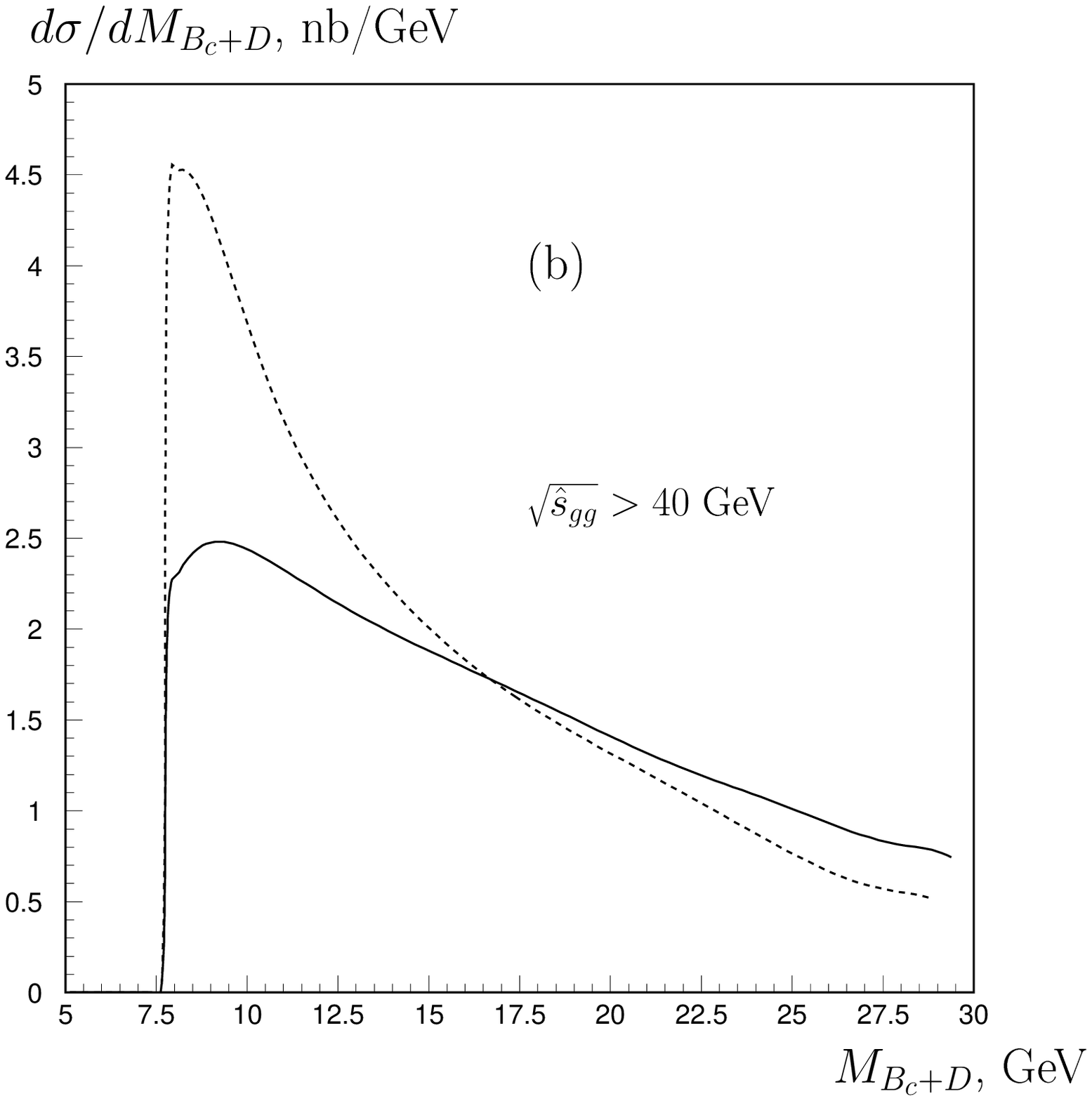}
}
\\
\parbox[t]{1.\textwidth}{
\caption{The cross section distributions over the invariant mass of $B_c$ and $\bar D$ mesons calculated within pQCD  for the process $pp \to B_c +X$ at $\sqrt{s}=14$~TeV: without cuts (a) and for $\sqrt{\hat s}>40$~GeV (b). Solid curves: $D_{c\to D}= \delta(1-z)$; dashed curves: $D_{c\to D}= z^{2.2}(1-z)$.
\hfill}
\label{fig:BcDminv_pp}}
\end{figure}
The cross section distributions over $M_{B_c+\bar D}$  are shown in Fig.~\ref{fig:BcDminv_pp} for the different kinematical regions. One can see that for the total phase space the cross section
distribution looks like the obtained within the fragmentation approach, but the cut on interaction energy essentially transforms the distribution shape. It  become essentially wider, whereas within the  fragmentation approach  it  should remain the same. Thus one can conclude that the recombination contribution is dominant.

\section{The angle correlations and the decay length ratio in the associative production of $B_c$ and $D$ mesons.}

\begin{figure}[!t]
\centering
\resizebox*{1.\textwidth}{!}{
\includegraphics{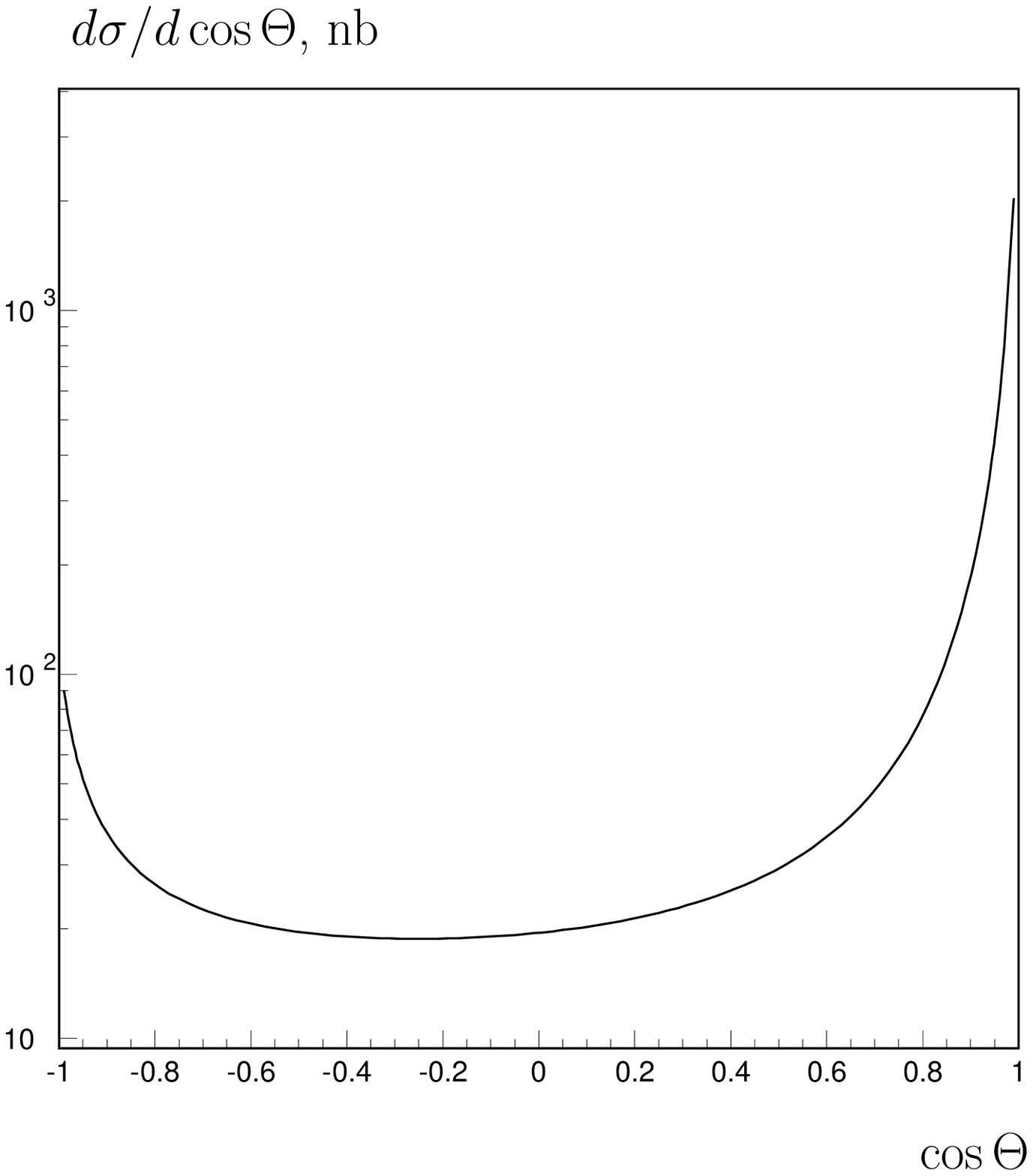}
\hfill
\includegraphics{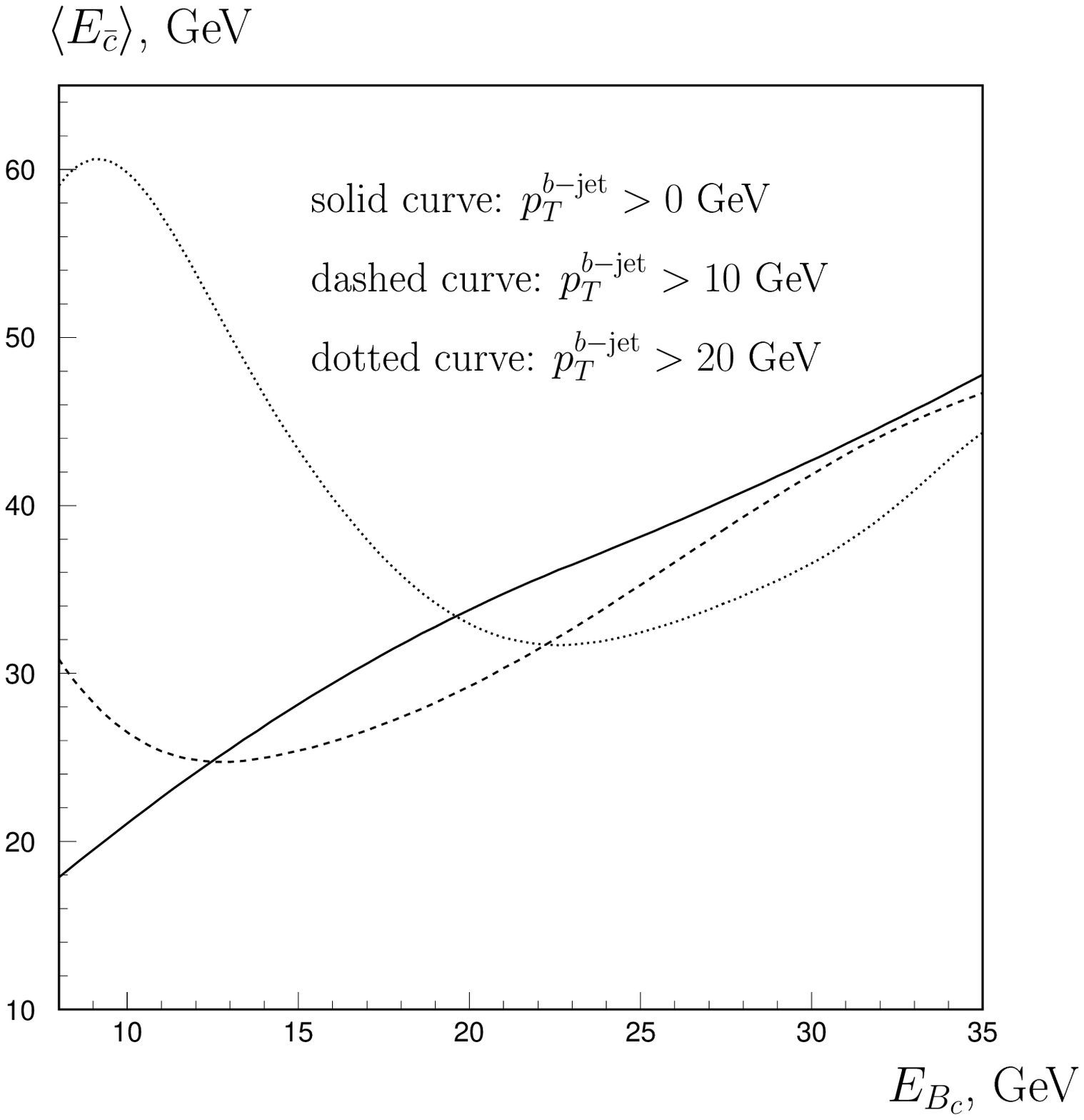}
}
\\
\parbox[t]{0.45\textwidth}{
\caption{The cross section distribution over the cosine  of angle between the  directions of motion of $B_c$ and $\bar D$ mesons predicted within pQCD  for the process $pp \to B_c +X$ at $\sqrt{s}=14$~TeV.
\hfill}
\label{fig:BcDcostheta_pp}}
\hfill
\parbox[t]{0.45\textwidth}{
\caption{The dependencies of averaged $\bar c$ quark energy on $B_c$ meson energy are represented for the different cuts on $b$-jet transverse momenta.}\label{fig:Bc_Ec_pp}}
\end{figure}

As it is shown in Fig.~\ref{fig:BcDcostheta_pp},  the cross section distribution on cosine of  the angle between the $B_c$ meson and $\bar D$ meson has the sharp maximum at
$\cos\theta\sim 1$. Moreover, about a half of $B_c$ mesons is associated by  the $\bar D$ meson moving in the close direction: $\theta \lesssim 26^o$.  Therefore, $\bar D$ meson can be used to detect $B_c$ meson.

In the Fig.~\ref{fig:Bc_Ec_pp} the dependencies of averaged $\bar c$ quark energy on $B_c$ meson energy are represented for the different cuts on $b$-jet transverse momenta.
It can be concluded from these plots that at any $b$-jet transverse
momentum 
\begin{equation}
\langle E_{\bar c} \rangle \gtrsim 1.2 E_{B_c}. 
\end{equation}

For $\bar D$ meson we obtain that 
\begin{equation}
\langle E_{\bar D} \rangle \gtrsim 0.7 \div 1.2 E_{B_c} 
\end{equation}
for $D_{c\to D}= z^{2.2}(1-z)$ and $D_{c\to D}= \delta(1-z)$,
correspondingly.

The decay lengths depend on particle energies and lifetimes as follows:
\begin{equation}
\langle l_{\bar D} \rangle \simeq \frac{\langle E_{\bar D} \rangle}{m_D} c \tau_D \qquad
 l_{B_c} \simeq \frac{ E_{B_c} }{m_{B_c}} c \tau_D.
\end{equation}

Taking into account that $\tau_D/\tau_{B_c}\simeq 2$ we obtain:
\begin{equation}
\frac{\langle l_{bar D} \rangle}{ l_{B_c} } \gtrsim 5.
\label{eq:length_ratio}
\end{equation}
This value should be compared with the fragmentation model prediction:
\begin{equation}
\frac{\langle l_{\bar D}^{\rm frag} \rangle}{l_{B_c}^{\rm frag} } \sim 1 \div 2.
\end{equation}

\section{Conclusions}

The following conclusion can be drawn from the performed calculations:

\begin{enumerate}
\item
The cross section distribution over the  invariant mass of $B_c$ and $\bar D$ meson   depends essentially on kinematical cuts and can be used to research $B_c$ production mechanism at LHC. 
\item 
In many cases the $B_c$ and $\bar D$ mesons move in close directions. It could be useful to detect $B_c$ meson.
\item
The energies of $B_c$ and $\bar D$ mesons are comparable. The decay length of $\bar D$ meson by more than $5$ times larger than the decay length of $B_c$ meson. The experimental research of the ratio  between $B_c$ meson and $\bar D$ meson energies could shed light on  $B_c$ production mechanism.
\end{enumerate}

\end{document}